# THE LOW LEVEL RF CONTROL SYSTEM OF PLS-II STORAGE RING AT 400 mA 3.0 GeV *

Inha Yu, Myunghwan Chun, Youngdo Joo, Insoo Park, Younguk Sohn, Mujin Lee, Sehwan Park, Seunghwan Shin PAL, San 31 Hyoja-dong, Nam-gu, Pohang 790-784, Korea

*Abstract*

The RF system for the Pohang Light Source (PLS) storage ring was greatly upgraded for PLS-II project of 400mA, 3.0GeV from 200mA, 2.5GeV. Three superconducting(SC) RF cavities with each 300kW maximum klystron amplifier were commissioned with electron beam in way of one by one during the last 3 years for beam current of 400mA to until March 2014. The RF system is designed to provide stable beam through precise RF phase and amplitude requirements to be less than 0.3% in amplitude and 0.3° in phase deviations. This paper describes the RF system configuration, design details and test results.

## INTRODUCTION

After three years of upgrading work, the Pohang Light Source-II (PLS-II) is now successfully operating. The final quantitative goal of PLS-II is a top-up user-service operation with beam current of 400mA to be completed by the end of 2014. Although available beam current is enhanced by setting a higher RF accelerating voltage, it is better to keep the RF accelerating voltage as low as possible in the long time top-up operation. We investigated the cause of the window vacuum pressure increment by studying the changes in the electric field distribution at the SC cavity and waveguide according to the beam current. The RF accelerating voltage of PLS-II RF system was set to 4.95 MV, which was estimated using the maximum available beam current that works as a function of RF voltage without stub tuners, and the top-up operation test with the beam current of 400mA was successfully.

**Table 1: Operation parameter of the PLS-II storage ring**

| Parameter | Unit | Values |
|---|---|---|
| Energy | GeV | 3 |
| Beam Current | mA | 400 |
| Circumference | M | 281.8 |
| Emittance | Nm-radian | 5.6 |
| RF acceptance | % | 2.8 |
| Accelerator Voltage, Vacc | MV | 4.5 |
| Energy loss per turn | keV | 1242 |
| Harmonic Number | | 470 |
| Momentum compaction factor | | $1.38 \times 10^{-3}$ |
| RF frequency | MHz | 499.973 |

The measured data set is shown in Fig. 1. During the test, the gap distance of IDs is set to the value of normal user service, and the total energy loss is about 1242 keV/turn (energy losses by radiation in bending magnets and IDs are 1042 keV/turn and 200 keV/turn, respectively). At first, the RF accelerating voltages of each SC are equally set to 1.6 MV as shown in the region A in Fig. 1. The SC2 RF accelerating voltage is controlled at slightly lower than the setting value because the LLRF output power limit is set to wrong value by mistake. Nevertheless, the stable top-up operation with beam current of 400mA beam is observed. The reflect power of SC2 is measured to be around 1–3 kW. In order to secure the operation margin and to reduce the chance of RF breakdown in the electric field at the RF window for a stable long time user-service operation, the RF accelerating voltages of each SC are equally set to 1.65 MV as shown in the region B in Fig. [1]

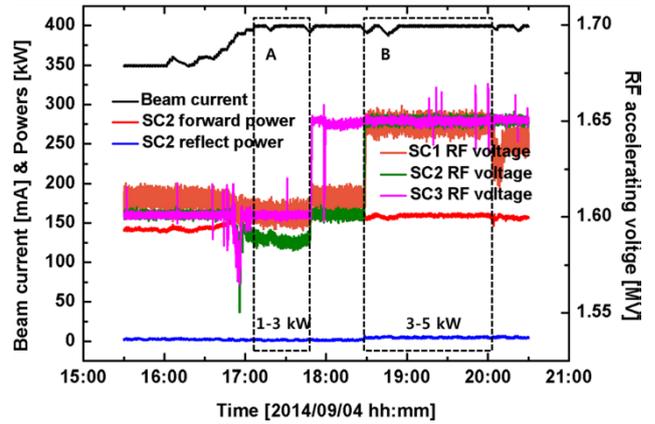

**Figure 1:** The beam current, RF voltages, forward power, and reflect power measured during the top-up operation test with the beam current of 400mA.

At this time, the LLRF power limit is set to right value and all the RF accelerating voltages are controlled at 1.5 MV. The reflection power is now increased to around 3–5 kW.

## CAVITY FIELD CONTROL RESULTS

**Table 2: Cavity field requirements and measured value**

| Electron Beam direction | Amplitude(rms) | | Phase(rms) | |
|---|---|---|---|---|
| | Spec. | Measured (@9-hr) | Spec. | Measured (@9-hr) |
| Cavity1 (1st) | < 0.3% | 0.18% (0.18) | < 0.3° | 0.2(0.21) |
| Cavity2 (2nd) | | 0.17% (0.17) | | 0.23(0.18) |
| Cavity3 (3rd) | | 0.18% (0.04) | | 0.24(0.26) |

The stability of the amplitude and phase of the cavity fields plays an important role in the beam energy spread, a critical figure of merit for nuclear physics experiments. Table 2 shows cavity field maximum allowable errors and the measured values according to 4.6MHz measurement bandwidth.

With the loop closed, we optimized proportional and integral gain by minimizing amplitude and phase noises. The optimized power spectral density of phase noise signal and its integrated RMS value measured for the SC cavity with 400mA electron beam in the PLS-II storage ring. [2]

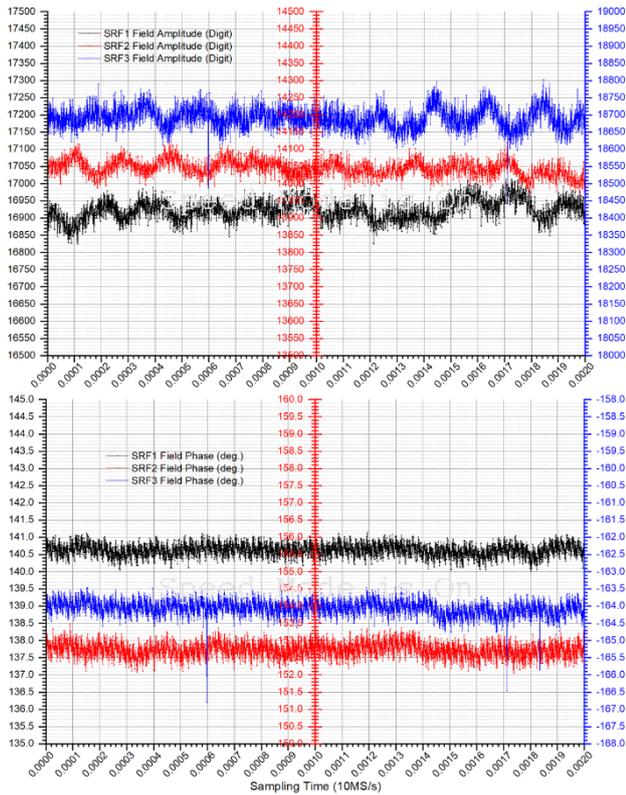

**Figure 2 : The cavity RF signal amplitude and phase variation during 2ms on 10MS/s, 5MHz measurement bandwidth**

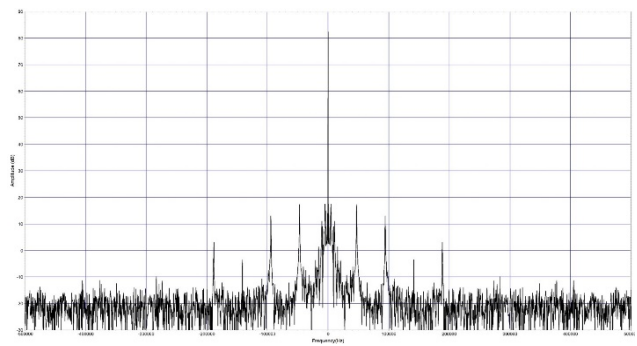

**Figure 3: The cavity RF sideband performance measured by 10MHz cycle ADC sampling**

Figure 2, 3 shows the amplitude, phase and RF carrier sideband performance were measured for three RF stations. Spurious signals were also observed less than -60 dB relative to RF carrier .

For 4.5MV (1.5MV/cavity) at 400mA at 3GeV in PLS-II storage ring the RF system performance was in the required control specification. Figure 4 shows the phase variation of the cavity fields at 3GeV 400mA beam during 9-hr.

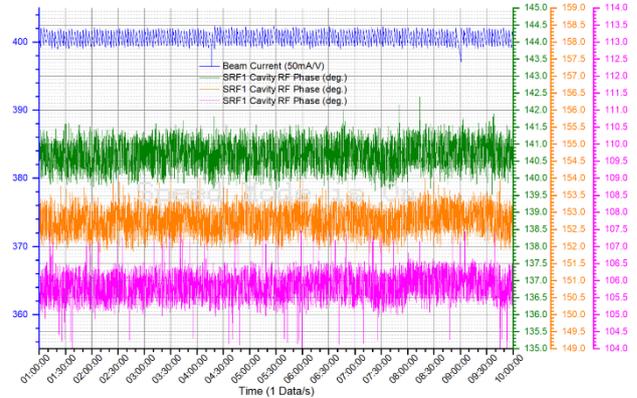

**Figure 4: SR RF cavities phase variation during at 3GeV 400mA beam**

## RF BLIP EFFECT SUPPRESSION

PLS-II SR third cavities suffer from blips on the RF probes as shown in Figure 5. It has very high amplitude. These blips may be due to electrons picked up by the probes. [3]

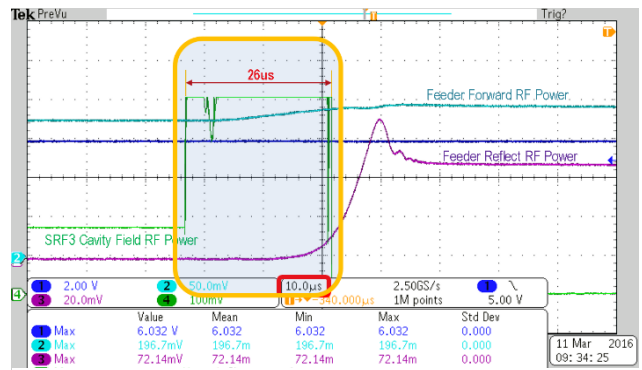

**Figure 5: RF blip signal from SRF3 cavity probe**

Figure 5 shows the post-mortem data using an oscilloscope by electron beam triggering. The third cavity signal fluctuated more than 26us with very highly amplitude with the reflected power increasing. The probe blips happen with beam losing. We are developing the controller for the cavity probe blip effect suppressing. When the blip happen the third cavity probe detected it is switch to the cavity's signal from the FBT near top (FBT) port to avoid a blip effect. The probe signal and the FBT signal phases is same proportionally but the FBT's amplitudes variation is larger as compared with the probe's signal amplitude and both signal phase is stable with constant proportion. We expect to avoid the beam losing from the blips of the third cavity. Figure 6 shows configuration of the controller for third cavity RF blip effect suppression. As shown Figure. 6. We are choses ADL5904 RF detector with rms power measuring and programmable envelope threshold detection function. The rms power measurement function has a 45 dB detection range. The envelope threshold detection function compares the voltage from an internal envelope detector with a user defined input voltage. When the voltage from the envelope detector exceeds the user defined threshold voltage, an internal comparator captures and latches the event to a set/reset (SR) flip-flop. The response time from the RF input signal exceeding the user programmed threshold to the output latching is 12 ns. The latched event is held on the flip-flop until a reset pulse is applied. [4] The controller with ADL5904 is designed and fabricated a 19" chassis device with the switched position keeping and reset times, threshold voltage (estimated RF blip level) setting functions.

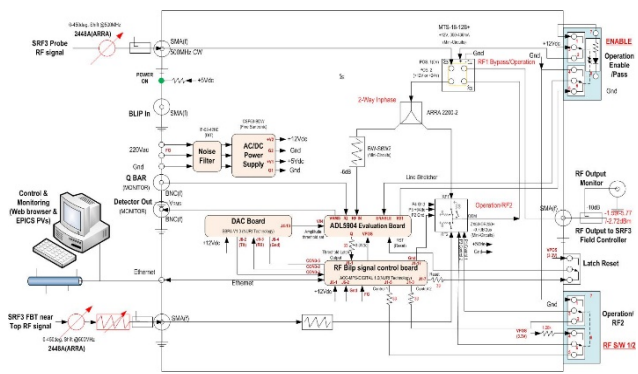

**Figure 6: The controller configuration for RF blip effect suppression**

## SR RF DIAGNOSTICS SYSTEM

A detailed description is shown fig. 7. The diagnostics system is including the related 32-CH 60MS/s digitizers Post-Mortem (4 8-CH PXI modules) in a NI LabVIEW application, 3 4-CH 100MHz bandwidth 2.5GS/s maximum oscilloscopes and the data storage devices with a 4-bay network attached storage solution equipped with a quad-core processor and 4GB DDR3L memory (Synology DS918+) and 4 20-TB Hard Disk Drives.

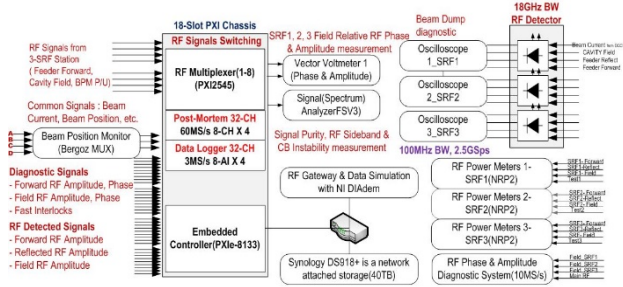

**Figure 7 : The configuration of the SR RF diagnostics with post-mortems and data simulation**

With Link Aggregation enabled, DS918+ delivers great sequential throughput performance at over 226 MB/s reading and 222 MB/s writing. The diagnostics system express the related before and after post-mortem acquisition data triggered by the SR beam current value and simulate in a LabVIEW program and NI DIAdem for finding and managing technical data, for mathematically and graphically-interactively analyzing data, and for presenting data in reports.

The data acquired from the variety of the different signals combined with the results of the beam current values from the beam current monitor (50mA/V), can serve to identify the cause of the beam trip. When it is an RF trip, it provides information regarding the signals immediately before and after the trip and accurate time stamping allows the sequence of events to be investigated.

The RF system trips can be classified according to their signatures. They may be classified into fast cavity vacuum trips, trips on the RF window, cavity quench, cavity arc and other trips. The most common and important trips in PLS-II are fast vacuum trips from the SC RF cavities. During such a trip, the cavity field collapses within a few microseconds and there are vacuum spikes on all the gauges around the cavity. [3] The PLS-II Storage Ring RF SC cavities also suffer from any probe problems same as DLS SR cavities probe problems. High amplitude fluctuated signals were observed just on the third cavity probes. These fluctuated signal events can confuse the LLRF leading to subsequent reaction, which may cause a beam trip.

## SUMMARY

The RF accelerating voltage of PLS-II RF system was set to 4.5 MV(1.5MV/cavity) with stub tuners, which have been operate well using the available beam current that works as a function of RF voltage, and the top-up operation with the beam current of 400mA. The RF system performance was satisfied the required control specification. We will try to raise the operation availability as decrease beam losing probability using the controller for third probe blip effect avoiding

## ACKNOWLEDGMENT

We would like to deeply thank P. Gu in DLS who gave much help and information. The Korea Ministry of Science & ICT (MSIP) supported this research.